\documentclass[aps,reprint,amsmath,notitlepage,prd]{revtex4-1}

\usepackage{amsmath} 
\usepackage{amssymb}
\usepackage{amsfonts}
\usepackage{amstext}
\usepackage{graphicx}
\usepackage{bm}
\usepackage{color} 
\usepackage{transparent}
\usepackage{mathtools}
\usepackage{slashed}
\usepackage{hyperref}

\newcommand{\dal}{\Box \phi}
\newcommand{\dpp}{\left(\nabla \phi \right)^2}

\renewcommand{\a}{\alpha}
\renewcommand{\b}{\beta}

\begin{document}

\title{Gravity with a generalized conformal scalar field: Theory and solutions}

\author{Pedro G. S. Fernandes}
\email{p.g.s.fernandes@qmul.ac.uk}
\affiliation{School of Physics and Astronomy, Queen Mary University of London, Mile End Road, London, E1 4NS, UK}

%\date{\today} 

\begin{abstract}
    We naturally extend the theory of gravity with a conformally coupled scalar field by only requiring conformal invariance of the scalar field equation of motion and not of the action. The classically extended theory incorporates a scalar-Gauss-Bonnet sector and has second-order equations of motion, belonging to the Horndeski class. Remarkably, the theory features a purely geometrical field equation that allows for closed-form black hole solutions and cosmologies to be easily found. These solutions permit investigations of in-vogue scalar-Gauss-Bonnet corrections to the gravitational action without the need of resorting to approximations or numerical methods. We discuss on the connection to the recently formulated 4D Einstein-Gauss-Bonnet theory of gravity.
\end{abstract}

\maketitle

%----------------------------------------------------------------------------------------------------
%\noindent{{\bf{\em Introduction.}}}
\section{Introduction}
%----------------------------------------------------------------------------------------------------
\par Einstein's theory of General Relativity is the most successful theory of gravity we have, predicting and explaining a plethora of observations. At the same time there are strong conceptual (e.g., quantization of gravity) and observational (e.g., puzzles in modern cosmology) reasons to expect that it should be modified, motivating the study of modified theories of gravity.

\par Lovelock theories of gravity \cite{Lovelock_original} are of particular interest because they are the only Lagrangian-based theories of gravity that give conserved, second-order, covariant field equations in terms of the metric only. In this sense, they are the most natural possible generalizations of General Relativity. The first of such non-trivial extensions of Einstein's theory occurs if the spacetime is five dimensional, where the Gauss-Bonnet term
\begin{equation}
\mathcal{G} = R^2-4R_{\mu \nu}R^{\mu \nu}+R_{\mu \nu \rho \sigma}R^{\mu \nu \rho \sigma},
\label{eq:GB}
\end{equation}
produces a rich phenomenology. In four dimensions, however, the Gauss-Bonnet term alone is known to be non-dynamical by virtue of Chern's theorem \cite{cherntheorem}. The consequences of making such a quadratic curvature correction to the gravitational action are often studied by invoking a new fundamental scalar field with a canonical kinetic term and coupling it to the Gauss-Bonnet term \cite{Sotiriou:2013qea, Sotiriou:2014pfa,Kanti:1995vq, Kleihaus:2011tg,Doneva:2017bvd, Silva:2017uqg, Antoniou:2017acq,Dima:2020yac,Herdeiro:2020wei,Berti:2020kgk} as motivated, for instance, by low-energy effective actions resulting from string theory, e.g. Einstein-dilaton-Gauss-Bonnet models \cite{zwiebach1985curvature,Kanti:1995vq,Kanti:1997br,Cunha:2016wzk}. In fact, due to Lovelock's theorem, modified theories in four dimensions will in virtually all cases involve additional degrees of freedom, usually regarded as new fundamental fields.

%\par Recent gravitational wave \cite{Abbott:2016blz,LIGOScientific:2018jsj} and black hole shadow \cite{Akiyama:2019cqa} observations are allowing us to test our theory of gravity and probe the structure of black holes with unprecedented accuracy. In electrovacuum, black holes are remarkably simple objects, being described by merely three macroscopic degrees of freedom (mass, electromagnetic charges, and angular momentum) and no other quantities colloquially referred to as \textit{hair}. While this simplicity is backed up by no-hair theorems \cite{NoHairChase,Hawking:1972qk,PhysRevD.51.R6608,Sotiriou:2011dz,Hui:2012qt,Herdeiro:2015waa}, new fundamental fields can lead to intriguing phenomenology in black hole spacetimes and if detected, may provide an insight into a manifestation of a more fundamental theory.

\par Closed-form solutions of the gravitational field equations allow for a simple inspection of the spacetime and calculation of observable predictions. However, modified theories with new fundamental fields typically present field equations with increased complexity such that these calculations become analytically impossible. One is then forced to resort either to perturbation theory, which is not well-justified in the extreme gravity regime, or to challenging numerical techniques \cite{Sullivan:2019vyi}.

\par Nonetheless, the Einstein equations with a matter source possessing conformal invariance are greatly simplified since the theory has constant scalar curvature on-shell, restricting the possible spacetimes and allowing for closed-form solutions to be easily found. An example of a theory with conformally invariant matter sources leading to simple solutions is precisely electrovacuum, whose Reissner-Nordstr\"om (Kerr-Newman) solution was the first ever discovered static (spinning) black hole with a matter source. One more example is gravity with a conformally coupled scalar field, whose matter action enjoys conformal invariance and is of the well-known form
\begin{equation}
\int d^4x \sqrt{-g} \left(\frac{R}{6} \Phi^2 +\left( \nabla \Phi\right)^2\right).
\label{eq:confcoupledaction}
\end{equation}
The first counter-example to the no-hair theorems (see e.g. Ref. \cite{Herdeiro:2015waa} for a review) was found precisely as a solution of this theory, the much-debated static BBMB black hole \cite{Bocharova:1970skc,Bekenstein:1975ts,Bekenstein:1974sf}. Due to its compelling properties, gravity with a conformal scalar field and its solutions have been extensively studied throughout the years (see e.g. Refs. \cite{Martinez:2002ru,Martinez:2005di,Anabalon:2009qt,Padilla:2013jza,deHaro:2006ymc,Dotti:2007cp,Gunzig:2000yj,Oliva:2011np,Cisterna:2021xxq,Caceres:2020myr} and references therein).

\par We note that in all the above examples the Einstein-Hilbert term explicitly breaks the conformal invariance of the full theory -- a remnant of this symmetry is only observed in the matter field equations of motion, such as the Maxwell equations or the modified Klein-Gordon equation resulting from the action \eqref{eq:confcoupledaction}. This suggests that the previously mentioned simplification of the equations of motion might in fact be related to the conformal invariance of the matter field equations and not of the action. Then, extended theories with the same effective symmetries \footnote{In this context we define an effective symmetry as a symmetry that is observed in some (but not all) of the equations of motion. E.g., conformal symmetry of the matter field equation which is not present in the other field equations.} might exist if conformal invariance is required solely in the matter field equation and not necessarily in the matter action. As we show below resorting to the example of a scalar field, the conformally coupled theory presented in Eq. \eqref{eq:confcoupledaction} can be extended in a natural way by incorporating a scalar-Gauss-Bonnet sector while preserving all of its effective symmetries. The extended theory presents field equations with a remarkable simplification that allows for simple closed-form black hole solutions and cosmologies, providing a framework to capture the essence of (scalar-)Gauss-Bonnet quadratic corrections to gravity in four-dimensions with analytical studies.%scalar-Gauss-Bonnet quadratic corrections to the gravitational action have attracted a great deal of attention in recent years.

%\par Hereafter we consider modified gravity, and in particular our theory of gravity with a generalized conformal scalar field, as a classical field theory extension of General Relativity, thus not explicitely adopting an effective field theory (EFT) point of view. Our reason to do so is motivated e.g. by low-energy effective actions from string theory where couplings between scalar fields and Gauss-Bonnet terms routinely occur \cite{zwiebach1985curvature,Kanti:1995vq,Kanti:1997br,Cunha:2016wzk}.

%\par Hereafter we adopt the point of view that modifications to General Relativity can occur as classical field theory extensions, instead of an effective field theory (EFT) perspective. Our reasoning is motivated e.g. by low-energy effective actions from string theory where couplings between scalar fields and Gauss-Bonnet terms routinely occur \cite{zwiebach1985curvature,Kanti:1995vq,Kanti:1997br,Cunha:2016wzk}.

\par Hereafter we consider classical modifications to general relativity, motivated from a ``top-down" point of view e.g. by low-energy-effective actions from string theory (where scalar-Gauss-Bonnet terms routinely occur \cite{zwiebach1985curvature,Kanti:1995vq,Kanti:1997br,Cunha:2016wzk}) rather than a ``bottom-up" approach inspired by Wilsonian effective field theory (EFT).

\par The paper is organized as follows: In Section \ref{s2} we derive the most general subset of the Horndeski family of theories \cite{Horndeski:1974wa,HorndeskiReview} whose scalar field equation of motion is conformally invariant and discuss on its compelling features. We then obtain static black hole solutions of the theory in Section \ref{s3} and briefly discuss its cosmologies in Section \ref{s4}.  We review our results and conclude in Section \ref{s5}.
We work with units where $c=1$, and use notation $\square \equiv \nabla_\mu \nabla^\mu$ and $\dpp \equiv \nabla_\mu \phi \nabla^\mu \phi$, where $\nabla_\mu$ is the covariant derivative.

%----------------------------------------------------------------------------------------------------
%\noindent{{\bf{\em A conformally invariant scalar field equation.}}}
%\subsection*{A conformally invariant scalar field equation.}
\section{Gravity with a generalized conformal scalar field}
\label{s2}
%----------------------------------------------------------------------------------------------------
We often denote with a tilde quantities constructed from the conformal geometry
\begin{equation}
\tilde{g}_{\mu \nu} = e^{2\phi} g_{\mu \nu},
\label{eq:tildemetric}
\end{equation}
which transforms as a metric under diffeomorphisms and is conformally invariant, with conformal transformations acting as \footnote{The transformation of Eq. \eqref{eq:weyltransf} is in fact a Weyl transformation, but we refer to it as a conformal transformation to be consistent with typical terminology.}
\begin{equation}
g_{\mu \nu} \to e^{2\sigma} g_{\mu \nu}, \qquad \phi \to \phi-\sigma,
\label{eq:weyltransf}
\end{equation}
where $\sigma \equiv \sigma(x)$ depends on the spacetime point. For convenience, we work with exponential conformal factors.

\par A remarkable property holds for scalar-tensor theories with a conformally invariant scalar field equation. Consider the transformation of Eq. \eqref{eq:weyltransf} in its infinitesimal form, such that $\delta_\sigma g_{\mu \nu} = 2 \sigma g_{\mu \nu}$ and $\delta_\sigma \phi = -\sigma$, where $\delta_\sigma$ denotes the change under an infinitesimal conformal transformation. In this case, an action describing a theory that depends solely on the metric $g_{\mu \nu}$ and a scalar field $\phi$, $S[\phi,g]$, varies by an amount
\begin{equation}
\begin{aligned}
\delta_\sigma S &= \int d^4x \left(\frac{\delta S[\phi,g]}{\delta g_{\mu \nu}}\delta_\sigma g_{\mu \nu} + \frac{\delta S[\phi,g]}{\delta \phi} \delta_\sigma \phi\right) \\&= -\int d^4x \left(-2 g_{\mu \nu}\frac{\delta S[\phi,g]}{\delta g_{\mu \nu}} + \frac{\delta S[\phi,g]}{\delta \phi} \right) \sigma,
\end{aligned}
\label{eq:trace1}
\end{equation}
where we identify the first and second terms in brackets with the trace and the scalar field equations, respectively. Now, if the scalar field equation is conformally invariant, then $\delta_\sigma S$ should be independent of $\phi$, such that the transformed action contains exactly the same scalar field dependence as the original one, resulting in the same scalar field equation. Thus, the quantity in brackets inside Eq. \eqref{eq:trace1}
\begin{equation}
-2 g_{\mu \nu}\frac{\delta S[\phi,g]}{\delta g_{\mu \nu}} + \frac{\delta S[\phi,g]}{\delta \phi},
\label{eq:geomcomb}
\end{equation}
should be a purely geometric quantity constructed only out of the metric $g_{\mu \nu}$. In short, a theory whose scalar field equation is conformally invariant, and not the matter action necessarily, will possess a purely geometrical field equation given by the sum of the trace and scalar field equations. This equation can be, in principle, more general than a constant scalar curvature condition and, at the same time, restrict the allowed spacetimes possibly providing an easy path to find closed-form solutions.

\subsection{Deriving the theory}

\par Our goal now is to derive the most general scalar-tensor theory with second-order equations of motion and a conformally invariant scalar field equation. To that end, we note that the derivatives of $\tilde{g}_{\mu \nu}$ are conformally invariant and so should be the curvature scalars constructed from it. In fact, we remark that scalar quantities constructed solely from the tilded metric are the only conformally invariant scalar quantities that depend only of the metric $g_{\mu \nu}$ and the scalar field $\phi$. The proof is similar to the one used in Ref. \cite{Padilla:2013jza} to construct the most general conformally invariant scalar-tensor action in four-dimensions, and is outlined next.

\par Let $\mathcal{I}[\phi,g]$ denote a scalar quantity that depends on the scalar field and the metric $g_{\mu \nu}$. Under a conformal transformation we obtain $$\mathcal{I}[\phi, g]\to \mathcal{I}[\phi-\sigma, e^{2\sigma}g].$$ Imposing conformal invariance, and choosing $\sigma=\phi$, we obtain $\mathcal{I}[\phi, g] = \mathcal{I}[0, \tilde{g}]$. Thus, the only conformally invariant scalar quantities that depend only on the scalar field $\phi$ and the metric $g_{\mu \nu}$ are purely geometric scalar quantities built out of the tilded geometry given in Eq. \eqref{eq:tildemetric}, $\mathcal{I}[0, \tilde{g}]$.
Therefore, in order for a theory to have a conformally invariant scalar field equation, the relation
\begin{equation}
\frac{\delta S[\phi, g]}{\delta \phi} = \sqrt{-\tilde{g}} \mathcal{I}[0,\tilde{g}],
\label{eq:tracesf}
\end{equation}
should hold. This will be our starting point to derive the sought theory of gravity with a generalized conformal scalar field. The reader will note that $\mathcal{I}[0, \tilde{g}]$ is yet unspecified. We will come back to that in a minute.

\par We will integrate Eq. \eqref{eq:tracesf} to obtain an action functional whose scalar field variation leads to a conformally invariant equation. The procedure to do so is similar to the one to reconstruct a function of several variables from its partial derivatives and is outlined next, following Ref. \cite[Section~9.7]{soper2008classical} closely. First, choose a configuration $\phi_c$ for the scalar field (typically taken to be zero) and a path $\phi(\eta)$  with $0 \leq \eta \leq 1$, that begins at the reference point $\phi(0) = \phi_c$ and leads to the desired final point $\phi(1)=\phi$. If an action $S[\phi,g]$ exists, then the construction will give the same result for any choice of the path. It is convenient to take a straight line path
\begin{equation*}
\phi(\eta) = \eta \phi + (1-\eta)\phi_c.
\end{equation*}
Consider now the action $S[\phi(\eta),g]$ evaluated along the path. We take its derivative with respect to $\eta$ obtaining
\begin{equation*}
\frac{dS[\phi(\eta),g]}{d \eta} = \int d^4x \frac{\delta S[\phi(\eta),g]}{\delta \phi(\eta)} \frac{d\phi(\eta)}{d\eta}.
\end{equation*}
Integrating from $\eta=0$ to $\eta=1$ we can obtain the sought action up to an integration constant functional $S_c[\phi_c,g]$, and taking $\phi_c=0$, we obtain
\begin{equation}
S[\phi,g] =\int d^4 x \int_{0}^1 d\eta \,\frac{\delta S[\phi,g]}{\delta \phi}\bigg\rvert_{\phi\to \eta \phi} \, \phi + S_c[0,g].
\label{eq:effaction}
\end{equation}

\par Let us review our progress so far. Our goal is to derive the most general subset of the Horndeski family of theories whose scalar field equation of motion is conformally invariant. We have shown that such theory will necessarily contain a purely geometric field equation that may provide an easy path to obtain closed-form solutions. Next we demonstrated that a conformally invariant scalar field equation must obey Eq. \eqref{eq:tracesf}. Finally we outlined a procedure to obtain the action functional that describes the theory, starting from the conformally invariant scalar field equation \eqref{eq:tracesf}. Obtaining the action that describes the sought theory then amounts to computing and simplifying Eq. \eqref{eq:effaction}.%So far we have not specified which quantities $\mathcal{I}[0,\tilde{g}]$ are suitable for Eq. \eqref{eq:tracesf}.

\par To work out which quantities $\mathcal{I}[0,\tilde{g}]$ are suitable for Eq. \eqref{eq:tracesf}, leading to an action made out only of $\phi$ and $g_{\mu \nu}$ whose equations of motion are second-order, we note the following. The quantity $\sqrt{-\tilde{g}}\mathcal{I}[0,\tilde{g}]$, once expressed in terms of $g_{\mu \nu}$ and $\phi$ inside Eq. \eqref{eq:effaction}, reveals that the action contains a term of the form 
\begin{equation*}
S[\phi,g] \supseteq \int d^4x \sqrt{-g} \int_{0}^{1} d\eta \, e^{(4-2k)\eta \phi}\phi \mathcal{I}[0,g],
\end{equation*}
with $k$ a constant related to the power of $\mathcal{I}[0,g]$ on the curvature (see e.g. Refs. \cite{Carneiro:2004rt,Dabrowski:2008kx} for a review of useful conformal transformations). As a result, the theory will necessarily contain non-minimal couplings of the scalar field to the geometric quantity $\mathcal{I}[0,g]$. Then, making use of Horndeski's theorem \cite{Horndeski:1974wa,HorndeskiReview}, the only scalar geometric quantity that can enter Eq. \eqref{eq:tracesf} without spoiling the requirement of second-order field equations is a linear combination of the form
\begin{equation}
\mathcal{I}[0,\tilde{g}] = -8\lambda - 2\beta \tilde{R} - \alpha \tilde{\mathcal{G}},
\label{eq:linearcombI}
\end{equation}
where $\lambda$, $\beta$ and $\alpha$ are constants and $\mathcal{G}$ is the Gauss-Bonnet term defined in Eq. \eqref{eq:GB}. We stress that we are interested only in theories whose equations of motion are of second-order to avoid Ostrogradsky instabilities in the classical theory, thus neglecting couplings to other curvature terms that, from an EFT point of view, are expected to be in the action on general grounds.

\par We will derive the effective action associated with each of the constituents of $\mathcal{I}[0,\tilde{g}]$ (i.e., the action whose scalar field variation leads to the respective constituent). To that end, given the conformal geometry of Eq. \eqref{eq:tildemetric}, we have the following useful relations in four dimensions \cite{Carneiro:2004rt,Dabrowski:2008kx}
\begin{widetext}
\begin{equation}
\begin{aligned}
&\sqrt{-\tilde{g}} = \sqrt{-g} e^{4\phi},\\&
\tilde{R} = e^{-2\phi} \left(R - 6 \dal -6 \dpp \right),\\&
\tilde{\mathcal{G}} = e^{-4\phi}\left[\mathcal{G}-8 R^{\mu \nu}\nabla_{\mu} \phi \nabla_{\nu} \phi+8G^{\mu \nu}\nabla_{\mu} \nabla_{\nu} \phi+8\dal\dpp-8\left(\nabla_{\mu} \nabla_{\nu} \phi\right)^2+8\left(\dal\right)^2+16\left(\nabla_{\mu} \phi \nabla_{\nu} \phi\right)\left(\nabla^{\mu} \nabla^{\nu} \phi\right)\right].
\end{aligned}
\label{eq:conformalscalars}
\end{equation}
In what follows, we resort to integration by parts, discard boundary terms and in the last steps absorb purely geometric terms built out of $g_{\mu \nu}$ into $S_c[0,g]$, as they have no effect in the scalar field equation. Starting with the effective action associated with $\lambda$ we use Eq. \eqref{eq:effaction} obtaining,
\begin{equation}
\begin{split}
S_{\lambda}[\phi,g] &= -8\lambda \int d^4x \sqrt{-g} \int_0^1 d\eta\ e^{4\phi \eta}\phi+S_c[0,g]
=-2\lambda\int d^4x \sqrt{-g} e^{4\phi}.
\end{split}
\end{equation}
Following the same procedure for the part of the total action associated with the conformal Ricci scalar
\begin{equation}
\begin{split}
S_{\tilde{R}}[\phi,g] &= -2\beta \int d^4x \sqrt{-g} \int_0^1 d\eta \,\, e^{2\phi \eta}\left(R-6 \eta \dal -6 \eta^2\dpp\right)\phi+S_c[0,g]\\&
=-2\beta \int d^4x \sqrt{-g} \bigg( \frac{e^{2\phi}R-R}{2}-3\dal \left(e^{2\phi}+\frac{1-e^{2\phi}}{2\phi}\right)-\frac{3\dpp}{2\phi^2} \left(e^{2\phi}-2\phi e^{2\phi}+2\phi^2 e^{2\phi} -1\right) \bigg)+S_c[0,g]\\&
=-\beta\int d^4x \sqrt{-g} e^{2\phi} \left(R+6 \dpp \right).
\end{split}
\end{equation}
The process to obtain the action associated with the Gauss-Bonnet term is similar and the following relations might prove useful:
\begin{equation*}
\begin{aligned}
&\nabla_\mu \left(\dal \nabla^\mu \phi - \frac{1}{2} \nabla^\mu \dpp\right) = (\dal)^2 - (\nabla_\mu \nabla_\nu \phi)^2 - R^{\mu \nu} \nabla_\mu \phi \nabla_\nu \phi,\\&
\int d^4 x \sqrt{-g} \phi \nabla_\mu \phi \nabla_\nu \phi \nabla^\mu \nabla^\nu \phi = -\frac{1}{2}\int d^4x \sqrt{-g} \left( \left(\nabla \phi\right)^4+\phi \dal \dpp \right) + \mbox{boundary terms}.
\end{aligned}
\end{equation*}
Then we obtain for the Gauss-Bonnet-related part of the action
\begin{equation}
\begin{aligned}
S_{\tilde{\mathcal{G}}}[\phi,g] &= -\alpha \int d^4x \sqrt{-g} \int_0^1 d\eta \bigg[\mathcal{G}-8 \eta^2 R^{\mu \nu}\nabla_{\mu} \phi \nabla_{\nu} \phi+8\eta G^{\mu \nu}\nabla_{\mu} \nabla_{\nu} \phi+8\eta^3\dal\dpp - 8\eta^2\left(\nabla_{\mu} \nabla_{\nu} \phi\right)^2\\&\,\,\,\,\,\,\,\,+8\eta^2 \left(\dal\right)^2+16\eta^3 \left(\nabla_{\mu} \phi \nabla_{\nu} \phi\right)\left(\nabla^{\mu} \nabla^{\nu} \phi\right) \bigg]\phi+S_c[0,g]\\&
= -\alpha \int d^4x \sqrt{-g} \bigg[\phi \mathcal{G}+4\phi G^{\mu \nu} \nabla_\mu \nabla_\nu \phi+2\phi \dal \dpp +4\phi \nabla_\mu \phi \nabla_\nu \phi \nabla^\mu \nabla^\nu \phi\\& \,\,\,\,\,\,\,\, +\frac{8}{3}\phi \big((\dal)^2 - (\nabla_\mu \nabla_\nu \phi)^2- R^{\mu \nu} \nabla_\mu \phi \nabla_\nu \phi\big)\bigg]+S_c[0,g]\\&
= -\alpha \int d^4x \sqrt{-g} \bigg[\phi \mathcal{G} - 4G^{\mu \nu} \nabla_\mu \phi \nabla_\nu \phi - 4\dal \dpp - 2(\nabla \phi)^4 \bigg].
\end{aligned}
\end{equation}
The final combined action can be obtained by summing all contributions, together with the Einstein-Hilbert term with a cosmological constant $\Lambda$ that does not affect the scalar field equation
\begin{equation}
\begin{aligned}
S&=\frac{1}{16\pi G} \left[ \int d^4x \sqrt{-g} \left(R-2\Lambda\right)+ S_\lambda + S_{\tilde{R}} + S_{\tilde{\mathcal{G}}}\right]\\
&=\int \frac{d^{4} x \sqrt{-g}}{16\pi G}\bigg[R-2\Lambda -\beta e^{2\phi}\left(R + 6(\nabla \phi)^{2}\right)-2\lambda e^{4\phi} - \alpha \bigg(\phi \mathcal{G} - 4 G^{\mu \nu} \nabla_{\mu} \phi \nabla_{\nu} \phi - 4 \square \phi(\nabla \phi)^{2} - 2(\nabla \phi)^{4}\bigg)\bigg].
\end{aligned}
\label{eq:actionconfgeneral2}
\end{equation}
The action given above in Eq. \eqref{eq:actionconfgeneral2} describes, up to field redefinitions, the most general scalar-tensor theory whose scalar field variation leads to a conformally invariant equation. It belongs to the Horndeski class with functions \footnote{Here we marginalized over the $1/16\pi G$ overall factor in the action.}
\begin{equation}
\begin{aligned}
G_2 = -2\Lambda -2 \lambda e^{4\phi} + 12\beta e^{2\phi} X + 8\alpha X^2, \quad G_3 = 8\alpha X, \quad
G_4 = 1-\beta e^{2\phi}+4\alpha X, \quad G_5 = 4\alpha \log X,
\end{aligned}
\label{eq:horndeskifuncs}
\end{equation}
where $X=-\frac{1}{2}\dpp$. 
The field equations are obtained by varying with respect to the metric
\begin{equation}
G_{\mu \nu} + \Lambda g_{\mu \nu} = -\alpha \mathcal{H}_{\mu \nu} + \beta e^{2\phi} \mathcal{A}_{\mu \nu}-\lambda e^{4\phi}g_{\mu \nu},
\label{feqs}
\end{equation}
where
\begin{equation*}
\begin{aligned}
\mathcal{H}_{\mu\nu} =& 2G_{\mu \nu} \dpp+4P_{\mu \alpha \nu \beta}\left(\nabla^\alpha \phi \nabla^\beta \phi - \nabla^\beta \nabla^\alpha \phi \right) +4\left(\nabla_\a \phi \nabla_\mu \phi - \nabla_\alpha \nabla_\mu \phi\right) \left(\nabla^\a \phi \nabla_\nu \phi - \nabla^\a \nabla_\nu \phi\right)\\
&+4\left(\nabla_\mu \phi \nabla_\nu \phi - \nabla_\nu \nabla_\mu \phi\right) \dal +g_{\mu \nu} \Big(2\left(\dal\right)^2 - \left( \nabla \phi\right)^4 + 2\nabla_\b \nabla_\a\phi\left(2\nabla^\a \phi \nabla^\b \phi - \nabla^\b \nabla^\a \phi \right) \Big),
\end{aligned}
\end{equation*}
\begin{equation*}
\begin{aligned}
\mathcal{A}_{\mu\nu} =& G_{\mu \nu} + 2\nabla_\mu \phi \nabla_\nu \phi - 2\nabla_\mu \nabla_\nu \phi +g_{\mu \nu} \left( 2\dal + \dpp \right),
\end{aligned}
\end{equation*}
with
$$
P_{\alpha \beta \mu \nu} \equiv *R*_{\alpha \beta \mu \nu} = -R_{\alpha \beta \mu \nu}-g_{\alpha \nu} R_{\beta \mu}+g_{\alpha \mu} R_{\beta \nu}-g_{\beta \mu} R_{\alpha \nu}+g_{\beta \nu} R_{\alpha \mu}-\frac{1}{2}\left(g_{\alpha \mu} g_{\beta \nu}-g_{\alpha \nu} g_{\beta \mu}\right) R,
$$
the double dual of the Riemann tensor, and the scalar field equation resulting from the action \eqref{eq:actionconfgeneral2} is equivalent to the vanishing of the quantity presented in Eq. \eqref{eq:linearcombI}
\begin{equation}
\beta \tilde{R}+\frac{\alpha}{2} \tilde{\mathcal{G}} + 4\lambda = 0,
\label{eq:sfeq}
\end{equation}
where the tilded quantities are defined in Eq. \eqref{eq:conformalscalars} in terms of $g_{\mu \nu}$ and $\phi$. Interestingly, the purely geometric combination \eqref{eq:geomcomb} results in the condition
\begin{equation}
R+\frac{\alpha}{2} \mathcal{G}-4\Lambda = 0,
\label{eq:traceeq}
\end{equation}
which is very similar to the trace equation of the higher-dimensional Einstein-Gauss-Bonnet theory.
\par The action described in Eq. \eqref{eq:actionconfgeneral2} can be cast into a more familiar form via the field redefinition $\Phi=e^{\phi}$
\begin{equation}
\begin{aligned}
S=\int \frac{d^{4} x \sqrt{-g}}{16\pi G}\bigg[R-2\Lambda - 6\beta\left(\frac{R}{6}\Phi^2 + \left(\nabla \Phi\right)^2\right)-2\lambda \Phi^4 - \alpha \bigg(\log(\Phi) \mathcal{G} - \frac{4G^{\mu \nu}\nabla_{\mu} \Phi \nabla_{\nu} \Phi}{\Phi^2} - \frac{4\square \Phi(\nabla \Phi)^{2}}{\Phi^3} + \frac{2(\nabla \Phi)^{4}}{\Phi^4}\bigg)\bigg],
\label{eq:actionconfgeneral}
\end{aligned}
\end{equation}
where we note the emergence of the usual conformally coupled scalar field action \eqref{eq:confcoupledaction} with a conformally invariant quartic potential. Observe that the action is invariant under the $\mathbb{Z}_2$ symmetry $\Phi \to -\Phi$.
\end{widetext}

%\section{Static black hole solutions}
%----------------------------------------------------------------------------------------------------
%\noindent{{\bf{\em Static black holes.}}}
\section{Static black hole solutions}
\label{s3}
%----------------------------------------------------------------------------------------------------
In this section we seek to obtain black hole solutions of the theory given in Eq. \eqref{eq:actionconfgeneral}, where we employ the static and spherically symmetric line element
\begin{equation}
ds^2 = -f(r) \,dt^2 + \frac{dr^2}{f(r)} + r^2 (d\theta^2 + \sin^2 \theta \, d\varphi^2 ).
\label{eq:ds2}
\end{equation}
If $\alpha=0$, the known solutions of the usual conformally coupled theory, such as the BBMB black hole, can be obtained. We are interested in the non-vanishing $\alpha$ case (furthermore, we assume $\beta \neq 0$). For the sake of completeness, we supplement the theory with the Maxwell action
\begin{equation}
S_{EM}=-\frac{1}{4} \int d^4x \sqrt{-g} F^{\mu \nu}F_{\mu \nu},
\label{eq:maxwellaction}
\end{equation}
where $F_{\mu \nu}=\partial_\mu A_\nu - \partial_\nu A_\mu$ is the Maxwell tensor and the subscript ``EM" stands for \textit{electromagnetic}. The associated (traceless) stress-energy tensor $T^{(EM)}_{\mu \nu}$ is
\begin{equation}
\begin{aligned}
T^{(EM)}_{\mu \nu} = F_{\mu \sigma} F_{\nu}^{\, \, \,\sigma}-\frac{1}{4}g_{\mu \nu} F_{\rho \sigma}F^{\rho \sigma},
\end{aligned}
\end{equation}
while the Maxwell equations are
\begin{equation}
\nabla_\mu F^{\mu \nu} = 0.
\end{equation}
Given the line element of Eq. \eqref{eq:ds2}, we assume a four-potential
\begin{equation}
A = V(r)\, dt - \frac{Q_m}{4\pi} \cos \theta \,d\varphi,
\end{equation}
with $Q_m$ the magnetic charge. The Maxwell equations imply
\begin{equation*}
V(r) = -\frac{Q_e}{4\pi r} - \Psi_e,
\end{equation*}
with $Q_e$ the electric charge and $\Psi_e$ the electrostatic potential. For future convenience we define
\begin{equation}
\mathcal{Q}^2=\frac{Q_e^2+Q_m^2}{4\pi}.
\label{eq:reducedcharge}
\end{equation}

\par Because any valid black hole solution must solve the geometric equation given in Eq. \eqref{eq:traceeq}, that for the line element \eqref{eq:ds2} takes the remarkably simple form
\begin{equation}
r^{-2}\left[\left(1-f\right) \left(r^2+\alpha (1-f) \right)\right]''-4\Lambda=0,
\end{equation}
with the prime denoting a radial derivative, a solution is easily integrated to be of the type
\begin{equation}
f(r) = 1+\frac{r^2}{2\alpha} \left[1\pm\sqrt{1+4\alpha \left(\frac{2 G M}{r^3}-\frac{q}{r^4}+\frac{\Lambda}{3}\right)} \right],
\label{eq:metricsolution}
\end{equation}
for any two integration constants $M$ (interpreted as the ADM mass) and $q$. The metric function with the plus sign before the square-root does not present a well-defined limit as $\alpha \to 0$, and has a non-physical asymptotic behavior near spatial infinity, so we disregard it as the physical one. A black hole described by the line element \eqref{eq:ds2} with $f(r)$ given in Eq. \eqref{eq:metricsolution} has, in the absence of the cosmological constant, horizons located at
\begin{equation}
r_\pm = G M \pm \sqrt{G^2M^2 - \alpha - q},
\end{equation}
and is asymptotically flat.

\par Now, a suitable linear combination of the $tt$ and $rr$ Einstein equations factorizes into a condition equivalent to
\begin{equation}
\left(\frac{\Phi'}{\Phi^2}\right)'\left(f \Phi' \left(r^2 \Phi\right)' + (f-1)\Phi^2-\frac{\beta}{2\alpha} r^2 \Phi^4\right)=0,
\end{equation}
that allows three distinct non-trivial branches
\begin{equation}
\begin{aligned}
&\mbox{(1)} \quad \Phi = \frac{c_1}{r+c_2},\\& \mbox{(2)} \quad \Phi = \frac{\sqrt{-2\alpha/\beta}\, \mbox{sech} \left(c_3 \pm \int^r \frac{dr}{r \sqrt{f}}\right)}{r},\\&
\mbox{(3)} \quad \Phi=c_4,
%\mbox{(3)} \quad \Phi=\frac{\exp\left(c_4\pm \int^r \frac{dr}{r \sqrt{f}} \right)}{r} \quad \mbox{if} \quad \beta=0,
\label{eq:scalarfieldprofile}
\end{aligned}
\end{equation}
where the $c_i$ are integration constants. We will analyze each situation in turn. The explicit equations of motion for the line element of Eq. \eqref{eq:ds2} are too long, intricate and not particularly elucidative to present here but can be found in the supplemental material \cite{SupplementalMaterial}.
\par For the first scalar field profile, the remaining field equations allow a static black hole solution with $f(r)$ given by Eq. \eqref{eq:metricsolution} with 
\begin{equation*}
q=G\mathcal{Q}^2-2\alpha, \quad c_1 = \sqrt{-2\alpha/\beta}, \quad c_2=0,
\end{equation*}
if $\lambda =\beta^2/4\alpha$, where $\mathcal{Q}$ is defined in Eq. \eqref{eq:reducedcharge}.
\par Assuming the second scalar field profile, the remaining field equations are solved with the metric function of Eq. \eqref{eq:metricsolution} provided that $q=G\mathcal{Q}^2$ as long as $\lambda=3\beta^2/4\alpha$. We note that the scalar field has a free parameter, $c_3$, that is not constrained by the field equations and so the scalar hair is, in a sense, primary. In all cases, the scalar field is regular on and outside the event horizon.

\par We remark that, in the absence of the electromagnetic field, a close inspection of the field equations \eqref{feqs} reveals the existence of a \textit{critical solution} with constant scalar field (third profile)
\begin{equation*}
\Phi=c_4=\sqrt{1/\beta},
\end{equation*}
as long as $\lambda=-\Lambda \beta^2$. In this extreme situation the Einstein equations \eqref{feqs} become an identity and we have left to solve only the purely geometrical condition of Eq. \eqref{eq:traceeq}, whose general solution is given by Eq. \eqref{eq:metricsolution} with unconstrained $q$.

\par A feature of the black holes here discussed is that they present an entropy $\mathcal{S}$, equal to the well-known Bekenstein-Hawking area term with a logarithmic correction
\begin{equation}
\mathcal{S} = \frac{A_+}{4G}+\frac{2\pi \alpha}{G} \log \left(\frac{A_+}{A_0}\right),
\end{equation}
where $A_+ = 4\pi r_+^2$ is the area of the event horizon and $A_0$ a constant with units of area. Logarithmic corrections to the black hole entropy often appear as the subleading term in several contexts, commonly related to quantum gravity \cite{Kaul:2000kf,Das:2001ic,Carlip:2000nv}.

%----------------------------------------------------------------------------------------------------
%\noindent{{\bf{\em FLRW Cosmology.}}}
\section{FLRW Cosmology}
\label{s4}
%----------------------------------------------------------------------------------------------------
In order to briefly study the cosmologies of the theory described by the action \eqref{eq:actionconfgeneral} we employ a flat FLRW background
\begin{equation}
ds^2 = -dt^2 + a(t)^2 d\mathbf{x}^2,
\label{eq:frw}
\end{equation}
and supplement the theory with matter content that has stress-energy tensor $T^{\mu}_{\,\,\,\, \nu} = \mbox{diag}\left(-\rho,p,p,p\right)$, where $\rho$ and $p$ are the energy density and pressure of the matter fields, respectively. We assume all quantities are homogeneous and isotropic functions of cosmological time. The explicit equations of motion for the spacetime of Eq. \eqref{eq:frw} are, again, too long and intricate to present here but can too be found in the supplemental material \cite{SupplementalMaterial}. 

\par It can be shown that the scalar field equation of motion \eqref{eq:traceeq} can be solved with a scalar field given by
\begin{equation}
%\Phi = \frac{2}{a} \left(c_1 \pm \sqrt{\frac{2\beta}{\alpha}\left(1\pm\sqrt{1-\frac{4 \alpha  \lambda }{3 \beta ^2}}\right)}\int^t \frac{dt}{a} \right)^{-1}.
%\Phi^{-1} = a \left(c_1 \pm \sqrt{\frac{\beta}{2\alpha}\left(1\pm\sqrt{1-\frac{4 \alpha  \lambda }{3 \beta ^2}}\right)}\int^t \frac{dt}{a} \right).
\Phi^{-1} = a \left(c_1 \pm \sqrt{\frac{\beta}{2\alpha}\left(\pm \sqrt{1-\frac{4 \alpha  \lambda }{3 \beta ^2}}-1\right)}\int^t \frac{dt}{a} \right).
\label{eq:sfcosmprofile}
\end{equation}
This profile, when substituted in the Einstein equations \eqref{feqs}, results in a set of modified Friedmann equations that take the remarkably simple form
\begin{equation}
\begin{aligned}
&\left(1+\alpha H^2\right)H^2=\frac{8\pi G}{3}\rho+\frac{\Lambda}{3},\\
&\left(1+2\alpha H^2\right)\dot H = -4\pi G (\rho + p),
\end{aligned}
\label{eq:friedmann}
\end{equation}
where $H=\dot a/a$ is the Hubble rate (the dot denotes a temporal derivative) and the matter fields obey the continuity equation $\dot \rho + 3H \left(\rho+p\right) = 0$. Remarkably, these cosmological equations have the same form as the ones that can be obtained in holographic cosmology \cite{Apostolopoulos:2008ru,Bilic:2015uol}, by invoking the generalized uncertainty principle \cite{Lidsey:2009xz}, and by considering a quantum corrected entropy-area relation of the apparent horizon of a FLRW universe \cite{Cai:2008ys}. Moreover they are equivalent to the ones resulting from the following set of field equations
\begin{equation*}
G_{\mu \nu} +\Lambda g_{\mu \nu} + \alpha\left(R_{\rho \mu \sigma \nu}R^{\rho \sigma}-\frac{1}{12} g_{\mu \nu} R^2\right)=8\pi G \,T_{\mu \nu}.
%G_{\mu \nu} +\Lambda g_{\mu \nu} + \alpha\left(\frac{1}{12} g_{\mu \nu} R^2-R_{\rho \mu \sigma \nu}R^{\rho \sigma}\right)=8\pi G \,T_{\mu \nu}.
\end{equation*}
The tensor in brackets on the left-hand-side has long been known to be covariantly conserved in conformally flat spacetimes on its own (it is sometimes called \textit{accidentally conserved}) \cite{ConsCFST,Mazur:2001aa,birrell1984quantum}, and appears naturally in the renormalized stress-energy tensor of matter fields in curved-spacetime \cite{birrell1984quantum}. %Moreover, it can be derived from a suitable dimensional regularization of the higher dimensional Gauss-Bonnet field equations to four-dimensions, assuming an identically vanishing Weyl tensor \cite{EGB31}.
%We remark that, even in the absence of both matter and the cosmological constant, de Sitter type solutions with $H=1/\sqrt{\alpha}$, are allowed by the modified Friedmann equations \eqref{eq:friedmann}.

\par Recently, the neutron star merging event GW170817 \cite{TheLIGOScientific:2017qsa} placed stringent constraints on the viable gravitational theories because the electromagnetic counterpart to GW170817 indicates that the deviation in the speed of gravitational waves, $c_T$, from that of light must be less than one part in $10^{15}$ \cite{Baker:2017hug}. The propagation speed of gravitational waves in Horndeski theories with non-trivial $G_4$ and $G_5$ functions differ, in general, from unity and is given by \cite{HorndeskiReview}
\begin{equation}
c_{T}^{2}=\frac{G_{4}-X\left(\ddot{\phi} G_{5, X}+G_{5, \phi}\right)}{G_{4}-2 X G_{4, X}-X\left(H \dot \phi G_{5, X}-G_{5, \phi}\right)}.
\label{eq:GWs}
\end{equation}
Consequently one might worry that the theory specified in Eq. \eqref{eq:horndeskifuncs} is severely constrained. Remarkably, however, a heavy constraint is easily evaded. If one considers a dark energy dominated universe where the scale factor is exponential $a \sim e^{\kappa t}$, using the scalar field profile of Eq. \eqref{eq:sfcosmprofile}, then $c_T^2=1$, leaving the theory unconstrained. Adopting a more conservative approach where we take $H^2 \approx -\dot H \approx 5.8 \times 10^{-36} \,\mbox{s}^{-2}$ and the fiducial value $\lambda=3\beta^2/4\alpha$, we obtain the constraint
\begin{equation}
\left|c_{T}^{2}-1\right| = \left| \frac{4\alpha \dot H}{1+2\alpha H^2}\right| \lesssim 10^{-15} \Rightarrow \sqrt{\left|\alpha \right|} \lesssim 10^{15} \,\mbox{km},
\end{equation}
which is a rather weak upper bound on $\alpha$.

%\section{Conclusions}
%\label{CONCLUSIONS}
%----------------------------------------------------------------------------------------------------
%\noindent{{\bf{\em Discussion.}}}
\section{Discussion}
\label{s5}
%----------------------------------------------------------------------------------------------------
Modified theories with new fundamental fields typically present field equations with increased complexity, forcing one to resort to perturbation theory or challenging numerical techniques. Nonetheless, the Einstein equations with a matter source whose action possesses conformal invariance, e.g. electrovacuum, are greatly simplified since the theory has constant Ricci scalar on-shell, restricting the possible spacetimes. This conformal symmetry is, however, only present in the matter field equations because the Einstein-Hilbert term spoils the conformal invariance of the full theory. In this work we have shown that if we solely require conformal invariance of the scalar field equation of motion and not necessarily of the action, then the conformally coupled scalar field theory of Eq. \eqref{eq:confcoupledaction} is naturally extended while preserving its effective symmetries and presenting field equations with a remarkable simplification.

\par We have computed the most general subset of Horndeski theories whose scalar field equation is conformally invariant. The theory is presented in Eq. \eqref{eq:actionconfgeneral} and is composed of the Einstein-Hilbert term, a cosmological constant, the action of the typical conformally coupled scalar field with a quartic potential, and a scalar-Gauss-Bonnet sector. It possesses a purely geometrical field equation that restricts the possible spacetimes, providing an easy path to find closed-form solutions. Several distinct static black hole solutions and a set of modified cosmological equations, meriting further investigation, were obtained in closed-form.

\par Gauss-Bonnet terms routinely occur in low-energy effective actions for gravity, and in particular show up as the leading-order correction in heterotic string theory \cite{zwiebach1985curvature, nepomechie1985low}. As previously stated in the introduction, in four dimensions, due to its topological character, the consequences of making such a quadratic curvature correction to the gravitational action are often studied by coupling a scalar field with a canonical kinetic term to a Gauss-Bonnet invariant through a coupling function as motivated by low-energy effective actions from string theory. These models have attracted a great deal of attention in recent years, where black holes in models with linear \cite{Sotiriou:2013qea, Sotiriou:2014pfa}, exponential \cite{Kanti:1995vq, Kleihaus:2011tg} and spontaneous scalarization-compatible \cite{Doneva:2017bvd, Silva:2017uqg, Antoniou:2017acq,Dima:2020yac,Herdeiro:2020wei,Berti:2020kgk} couplings have all been extensively studied as have their cosmologies \cite{
Nojiri:2005vv, Jiang:2013gza, Kanti:2015pda, Chakraborty:2018scm, Odintsov:2018zhw, Odintsov:2019clh, Odintsov:2020zkl}. The author is not aware of any known closed-form black hole solutions of the aforementioned models, thus the ones reported in this work serve as a good proxy to capture the essence and probe scalar-Gauss-Bonnet effects with analytic studies.

\par Interest in the dimensional regularization of gravity with a Gauss-Bonnet term was recently spurred by the original work of Ref. \cite{original}. Even though the original procedure presented in Ref. \cite{original} is not well-defined \cite{EGB31,Gurses:2020rxb}, similar (healthy) regularization procedures can be performed, resulting in well-defined four-dimensional theories that typically present a new scalar degree of freedom \cite{regularization,otherone,EGBST1,EGBST2,obs4degb} (also see \cite{Aoki:2020lig}), and remarkably constitute particular cases of the full theory given in Eq. \eqref{eq:actionconfgeneral2}. As such, these works stumbled into a purely geometric field equation similar to Eq. \eqref{eq:traceeq}, whose existence had no seemingly profound explanation. Our paper then reveals that the existence of such field equation is intimately connected with (generalized) conformal properties of the scalar field. It is rather intriguing that the dimensional regularization procedures of the Gauss-Bonnet term done in the works \cite{regularization,otherone,EGBST1,EGBST2,obs4degb} lead to particular cases of the theory derived in this paper. Further work is being done on investigations of possible fundamental (stringy) origins for these regularized theories \cite{Easson:2020bgk}. For other relevant works on these subsets of the full theory given in Eq. \eqref{eq:actionconfgeneral2} we refer the reader e.g. to the study of its vacua \cite{Ma:2020ufk}, observational constraints \cite{obs4degb}, tree-level graviton scattering amplitudes \cite{Bonifacio:2020vbk}, and studies related to lower-D limits of the regularization procedure \cite{Lu:2020mjp,Hennigar:2020fkv,Hennigar:2020drx}.

\par The part of the action \eqref{eq:actionconfgeneral2} associated with the Gauss-Bonnet term coincides precisely with the local effective action that one obtains for the trace anomaly that results from the broken conformal symmetry of massless fields in quantum theory \cite{Riegert:1984kt,Komargodski:2011vj}. Moreover, the non-minimal coupling of Eq. \eqref{eq:confcoupledaction} is sourced by quantum corrections even if set to zero in the classical action and it has been argued that it is naturally expected at high curvatures \cite{birrell1984quantum,Birrell:1979ip,Ford:1981xj,PhysRevD.35.2955}.
In a sense, gravity with a generalized conformal scalar field can be perceived as a gravitational theory that displays known quantum corrections.% Further support to this claim comes, e.g., from the logarithmic corrections to the entropy of the horizons.

\par Primordial black holes that might have formed with masses smaller than $\sim 10^{15}$ g are typically disregarded as dark matter candidates, since they would have evaporated by now \cite{Carr:2020gox}. The static black holes here reported, however, can in some circumstances present a minimum size, $r_+^{min} =\sqrt{|\alpha|}$ and evaporation would cease at this Gauss-Bonnet related scale, leaving relics \footnote{Here we neglected the cosmological constant and electromagnetic charges.}. If these black holes exist in Nature, and are sufficiently stable, they could in principle constitute a part of the dark matter, provided that $\alpha$ is not too big. It would be interesting to explore this hypothesis further.

\par Other possible avenues of research include e.g., performing a more detailed study of both the black holes and the cosmologies of the theory; exploring the existence of generalizations of the idea here presented to other types of matter, such as vector fields; study gravity with a generalized conformal scalar field in a number of dimensions different than four. For instance, in two dimensions the resulting generalized conformal scalar field theory is the well-studied Liouville theory of gravity \cite{Jackiw:2005su,Grumiller:2007wb}
\begin{equation*}
S_{2D} \sim \int d^2x \sqrt{-g} \left(-\Lambda + \phi R  + \dpp + \mu e^{2\phi}\right),
\end{equation*}
whose equation of motion is of the type $R+2\Lambda \sim T$, therefore behaving as Einstein's gravity \footnote{In two dimensions there is only one degree of freedom in the geometry, which means that the purely geometrical field equation contains all information about the theory.}. In six dimensions one could try to obtain non-trivial dynamics from the cubic Euler density, much like we did in this work with the quadratic Gauss-Bonnet term in four dimensions.

%%%%%%%%%%%%%%%%%%%%%%%%%%%  
%----------------------------------------------------------------------------------------------------
%\noindent{{\bf{\em Acknowledgements.}}}
\subsection*{Acknowledgements.}
%----------------------------------------------------------------------------------------------------
%%%%%%%%%%%%%%%%%%%%%%%%%%%
The author is supported by the Royal Society Grant No. RGF/EA/180022, acknowledges support from the CERN Project No. CERN/FISPAR/0027/2019, and thanks Bruno V. Bento, Pedro Carrilho, Timothy Clifton, Carlos Herdeiro, David Mulryne and Sofia Pinto for helpful comments and discussions. Calculations were verified using the Mathematica packages \textit{diffgeo.m} \cite{diffgeo} and \textit{GREAT.m} \cite{great}.

\bibliography{biblio}

\begin{thebibliography}{86}
\bibitem [{Sup()}]{SupplementalMaterial}%
  \BibitemOpen
  \href@noop {} {}\bibinfo {note} {{See ancillary file for a Mathematica
  notebook that computes explicitly the field equations for both an FLRW and a
  static, spherically symmetric metric ans\"atz.}}\BibitemShut {Stop}%
\end{thebibliography}

\end{document}